# Re-Imagining HCI: New Materialist Philosophy and Figurations as Tool for Design

**Goda Klumbytė**
Gender/Diversity in Informatics Systems
University of Kassel, 34121 Kassel, Germany
goda.klumbyte@uni-kassel.de

**Claude Draude**
Gender/Diversity in Informatics Systems
University of Kassel, 34121 Kassel, Germany
claude.draude@uni-kassel.de

**Loren Britton**
Gender/Diversity in Informatics Systems
University of Kassel, 34121 Kassel, Germany
loren.britton@uni-kassel.de

## ABSTRACT

In this paper[*] we interrogate the practices of imagining in human-computer interaction (HCI), particularly in scenario building (SBE) and persona construction. We discuss the philosophical premises of HCI imaginings in rationalism, cognitivism and phenomenology, and we propose (feminist) new materialist philosophy as an enriching perspective that helps generate a holistic, relational perspective of users, imaginaries and technologies. In the end we explore the method of figurations as a potential tool for HCI design.

## INTRODUCTION

In this paper we interrogate the practices of imagining in human-computer interaction (HCI), particularly in scenario building (SBE) and persona construction. Recognizing the importance of SBE and Personas as methods that help bring the user into focus, we propose a new materialist, particularly feminist new materialist philosophical framework and the method of figurations/figuring, as an enriching approach that can help extend user-orientation towards a more holistic perspective in HCI design that focuses on the relational and extended account of users and environments. The key questions driving this perspective are: Where do we imagine from and how are imaginaries built? What gets accounted for? How can imaginaries be constructed otherwise? And are there limits to what kind of imaginaries are usable for computing?

To interrogate these questions, we start with a brief overview of SBE and persona construction in HCI. Then we discuss several approaches to imaginaries and their philosophical underpinnings, particularly rationalism, phenomenology and cognitivism. We then introduce new materialist philosophy and figurations as a methodological tool. We close with some suggestions of how new materialist approach can benefit HCI research and design practices.

**SCENARIO BASED ENGINEERING AND PERSONAS: MODELS AND CRITIQUES**

Imaging practices and the construction of imaginaries is an important aspect of technology development and design. In computation, scenario-based engineering and construction of personas are commonplace methods for imagining the situation and context of the future technology performance and use (30,14, 4, 21). Through these practices certain scripts are developed that influence how a particular technological tool will be designed and how it is supposed to be used. Particularly SBE has been used widely to analyze how prospective technology might affect user activities and modes of interaction. SEB is sectioned into the sub-categories of problem scenarios, activity scenarios and information and interaction design scenarios. In all these scenarios certain actors or agents – imagined users based on specific roles – play an important part to identify and model modes of interaction in specific cases or problem fields (4).

Persona building is another common practice in HCI and relies on translating data as well as existing assumptions about users into fictious yet concrete representations (1). The advantage of personas is that they make implicit assumptions about the users explicit, as well as allow to go beyond the rather impersonal "user" by adding context and personality and thus allowing technology developers to better imagine the anticipated situations of interaction and understand the needs of the users. In a sense, they "materialize" the user through
giving them personal characteristics, interests and concrete situations of interaction. Such personas can also be incorporated into SBE in order to improve the relatability to actors in specific scenarios.

These practices show how crucial metaphors and imaginaries are for HCI design and development. Furthermore, they also perform a function of allowing the designers to relate to the users through empathy to actors that are personified enough to be depicted as "real human beings". Furthermore, both SBE and persona building are expected to be evidence-based, thus showing how selection of purely imaginary positions and situations as well as assumptions and potentially even stereotypes blend with factual information. Precisely because such practices blend fact and fiction, it is crucial to ask from which positions do these imaginaries and imagining practices stem from.

**Allegory of the Cave**

Allegory of the Cave is a thought experiment that depicts a cave where prisoners are chained facing a wall, unable to move their heads. Behind them burns a fire, in front of which, but invisible to prisoners, a puppet show takes place. The prisoners are unable to see the puppets, only the shadows on the wall that they are facing. Because the shadows are their primary experiential reality, the prisoners would be unable to talk of what "really" is beyond the shadows, unless they grasp the causes of the shadows. As they are unable to turn their heads, the only way they can grasp such true reality is mentally.

Source:

https://faculty.washington.edu/smcohen/320/cave.htm.

**PHILOSOPHICAL UNDERPINNINGS OF IMAGINING IN HCI**

What is at stake in imagining practices is not only the kind of ideas and imaginaries that are embodied in technology but also the premises of knowledge production through computational tools and practices in general. Such premises stem not least from philosophical traditions, including rationalism, phenomenology and cognitivism, all of which can provide different sets of starting points for HCI. Rationalism, in a most simplified sense, proposes that true knowledge relies on reason, as opposed to sense experience (28). A historical example of rationalism is Plato's allegory of the cave and his idea that only those truths that are grasped through reason are the real, unchanging ones, as opposed to sense experience that is always shifting (cf. also Descartes – 28). Cognitivism is in this respect related to rationalism because it relies on cognitive models to account for how people reason and make decisions. Different forms of cognitive can espouse rather different explanations of cognition and perception (e.g. the 4E cognitive approach takes cognition to be embodied, embedded, enactive and extended – 29).

Both rationalism and cognitivism are useful for HCI because they propose accounts of human behaviour and thought that are easier to formalize into clear design models and allows to capture generalities of human behaviour on an abstract level. However, this also makes them susceptible to universalism and thus they run the risk of blanketing crucial cultural and other differences. They also might prioritise mind over body and thus mental experiences versus sensory engagement, as well as miss the cultural and political significance of different embodied positions of users. Phenomenology, which is a philosophical tradition that specifically focusses on the body and the lived experience, remedies that through drawing attention back to the embodied experience (36), as well as how such experience changes depending on race, class, gender and sexuality (e.g. 2,3).

All of these three major traditions in their classical form have several points in common. First, they all rely on the human as a universal (even if possibly internally differentiated across gender and other lines) subject and object of knowledge. However, as many feminist, anti-racist, disability studies and trans*gender studies critiques have shown (11, 15, 23, 35), who gets to be included in the category of the human, has been historically contingent. Therefore, certain knowledges and perspectives that tend to claim universality might ignore how their own point of vision is specific and partial. Relatedly, these positions are also profoundly anthropocentric in that they take first and foremost the perspective and therefore interests of humans into account. This can be problematic specifically in the current predicament of wide-spread environmental devastation and 6[th] Extinction (12), which necessitates thinking about environment and non-human beings as stakeholders in human technology development. Lastly, these positions encourage the perspective of the user as an individual, atomistic subject, and such bounded individual is then taken as primary reference point and basic unit of analysis. Such perspective does not account for the profound embeddedness and relationality of human existence, subjectivity formation and development.

**Niels Bohr and Diffraction Patterns**

N. Bohr was studying how to resolve the question of whether light is a wave or a particle, since depending on how the apparatus of diffraction is set, it will either show the wave-like pattern of interference, suggesting that light acts as waves, or not, suggesting that rather light has particle-like characteristics. Bohr resolved this by proposing that the referent is not an independently existing entity but rather "the phenomenon of light intra-acting with the apparatus" (5).

**NEW MATERIALIST PHILOSOPHY**

In feminist theory, imaginings and imaginaries have played a prominent role in fields such as psychoanalysis, philosophy and science studies. Feminist scholars have explored phallo-logo-centric basis of psychoanalytic imaginaries (20, 13, 24, 25), the white, heterosexual male as the key imaginary for humanist theories of modern subjectivity and philosophy (8, 27), and the anthropocentric, gendered set of imaginaries as the starting point for technoscience (16, 22, 31, 32).

The emerging new materialist thought, and particularly feminist new materialism (FNM), brings to attention how imaginaries are materially embedded and embodied through focusing on the entanglements between matter and discourse, nature and culture, epistemology and ontology. FNM proposes to re-think three key concepts. First, it sees matter – biological, technological, organic and inorganic matter – as vital, self-organizing force. Second, it challenges the ontological distinctions that are dominant in Western philosophy, such as nature/culture, man/woman, human/non-human, self/other and proposes that entities co-define each other. Third, agency, and specifically also human agency and subjectivity, is re-defined as profoundly relational. This means that entities (including human subjects) do not precede relations but emerge in them.

FNM proposes to think, as Karen Barad suggests, ethico-onto-epistemologically, or ethics, being and knowing as inseparable (5). In her work on re-reading Bohr's quantum physics for feminist philosophy, Barad critiques the representationalist idea of subjectivity and objectivity, and the tripartite relation of the knower – the knowledge – and the known. Instead, she proposes something that she calls "agential realism". Agential realism is a stance that suggests that entities – knowledge, things, subjects – or phenomena are made and unmade through intra-action. This means that they are defined in relations and do not pre-exist them as separate, stable entities. According to Barad, it is through apparatuses, which themselves are material-discursive practices, that assemblages of matter become known as stable entities. These apparatuses perform "agential cuts" that stabilize matter into subjects and objects within intra-actions. A good example of such "agential cuts" would be different scientific traditions and their measurement tools

To summarize, FNM has profound philosophical implications for how technologies, humans, environments and their inter-relations are accounted for. Continuing the feminist tradition of relational ethics and subject formation, FNM posits relations before entities. By focusing on the agency of non-human entities, it also provides space for a more holistic and animated account of HCI environments and material agencies that technologies, practices, discourses as well as human subjects come to have in the co-defining intra-action. Lastly, it proposes that we cannot disentangle ways of seeing and knowing (epistemology) from ways of being (ontology) and thus encourage ethical responsibility for the apparatuses we use and the "agential cuts" that get made in the process.

**FIGURATIONS AS NEW MATERIALIST TOOL**

The notion of *figuration* has been crucial to feminist theorizing and politics, and specifically FNM. Rosi Braidotti defines figurations as *conceptual personae,* "materialistic mappings of situated, i.e. embedded and embodied, social positions" (10, p2). Figurations are material-discursive entities that account for particular historical, political and material locations. Donna Haraway in her work

**Haraway's Figuration: Cyborg**

One of the first of Haraway's own figurations developed in the 1980s was the cyborg (17), which she used both to account conceptually for the relationship between humans and technology, the power relations and biases implicated in such relation, as well as to point to modes of ethical responsibility in the production and use of technologies. Other subsequent figurations that Haraway thought with included: companion species, kin, chthulucene, genetically modified mouse, and others.

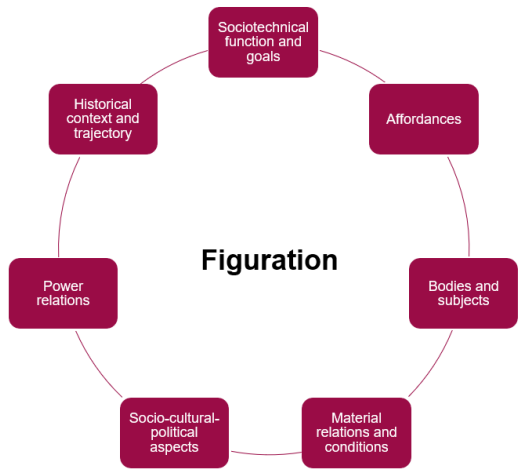

Figure 1. Figuration scheme by authors

highlights that figurations stitch together meanings and practices. In her line of thinking, technologies are materialised figurations that bring together both actual physical technologies and clusters of meaning (narratives, discourses, imaginaries) surrounding them, which together form more or less stable assemblages or *configurations* (34). Feminist phenomenologists such as Sara Ahmed also point out that the daily encounters between such configurations and bodies shape their interactions with the world: our (gendered) bodies and material objects take shape through being oriented towards each other (2). This mutual orientation and intra-action (5) is crucial for contemporary technofeminist practices and interventions (33).

The way FNM figurations are constructed is often through mapping or cartography (10). Cartography works as a special layering of material, theories and discourses on top of each other and mapping out significant points of connection. It is a way to generate a "theoretically-based and politically-informed reading[s] of the present" (9, p3) that allow to think simultaneously of several coexisting phenomena or conditions. Similarly, Donna Haraway proposes that figurations are built by creating "performative images that can be inhabited" (18, p11). This means that they are both materially grounded as well as engender the potential to trouble existing imaginaries and representations (a comparable method could be that of archaeological re-construction of the past or using archaeologically inspired methods for utopian imaginaries – cf. 6, 26). What is notable about this method is that the materials used for building figurations always include historical dimension, political dimension, accounts of power imbalances, embodied and embedded accounts of concerned subjects or subject groups, and the resulting figuration is not necessarily expected to take human form or character.

Figurations as a concept and a method could bring several benefits for HCI. First, it brings attention to the systemic level rather than only user-focus, and challenges researchers and designers to situate themselves, situate and imagine their subjects (users) as relational, embedded and contextualized. This systemic perspective also allows to expand the area of concern beyond immediate effects of or interaction with technology by the human users, but also incorporate ecological concerns and structural concerns (such as possible power dynamics). Conceptually thus it opens a space to challenge anthropocentrism even while we still design technologies for human users.

Further, figuration as a tool can be applied not simply in thinking about the user (since Personas might do justice to that already), but rather in thinking about systems themselves as material-discursive configurations. Focus on relationality through figurations also would allow to more explicitly re-focus on practices and relations. This could enable to ask the questions: what kind of relations does a system facilitate? Who benefit from such relations? What kind of agencies emerge in it? Importantly, figurations always include historical dimensions. This means that part of situating the users and researchers, as well as the system itself, questions of historical origins, heritages and unequal distributions of power are always considered (cf. 7, 19).

**CONCLUSION**

In this paper we analysed scenario-based engineering (SBE) and persona building as one of the key tools of imagining in HCI and their philosophical roots in rationalism, phenomenology and cognitivism. We then proposed a feminist new materialist philosophy (FNM) as an enriching lens that highlights the agency of matter (including technological objects), ontological relationality, entanglements of matter and discourse, and the inseparability of ontology and epistemology. We proposed that using an FNM method of figurations can help bring a more holistic and relational perspective to HCI design. We would like to conclude by noting that this is a highly experimental suggestion, and research into intersections between FNM and computing is still in its germinal state. The authors hope, however, that this experimental proposal can help open new perspectives in the intersections between HCI and philosophy and help advance HCI design methods.

## REFERENCES

[1] Tamara Adlin and John Pruitt. 2008. Putting Personas to Work: Using Data-Driven Personas to Focus Product Planning, Design and Development. In *The Handbook of Human Computer Interaction* (2nd ed.), eds. Andrew Sears and Julie A. Jacko. Lawrence Erlbaum Associates, New York, USA, London, UK, 991-1016.
[2] Sara Ahmed. 2006. *Queer Phenomenology*. Duke University Press, New York, USA.
[3] Sara Ahmed. 2007. A Phenomenology of Whiteness. *Feminist Theory*, 8(2), pp. 149-168.
[4] Ian F. Alexander, Neil Maiden. 2004. *Scenarios, Stories, Use Cases. Through the Systems Development Life-Cycle*. Wiley, New Jersey, USA.
[5] Karen Barad. 2007. *Meeting the Universe Halfway: Quantum Physics and the Entanglement of Matter and Meaning*. Duke University Press, Durham, USA.
[6] Shaowen Bardzell. 2018. Utopias of Participation: Feminism, Design, and the Futures. ACM Trans. Comput.-Hum. Interact. 25, 1, Article 6 (February 2018).DOI: https://doi.org/10.1145/3127359
[7] Shaowen Bardzell and Jeffrey Bardzell. 2011. Towards a feminist HCI methodology: Social science, feminism, and HCI. *Proceedings of CHI 2011*, ACM, New York, 675-684.
[8] Rosi Braidotti. 1991. *Patterns of Dissonance: A Study of Women in Contemporary Philosophy*. Polity Press, Cambridge, UK.
[9] Rosi Braidotti. 2002. *Metamorphoses: Towards a Materialist Theory of Becoming*. Polity Press, Cambridge, UK.
[10] Rosi Braidotti. 2011. *Nomadic Subjects: Embodiment and Sexuality in Contemporary Feminist Theory*. Columbia University Press, New York, USA:
[11] Rosi Braidotti. 2013. *The Posthuman*. Polity Press, Cambridge, MA, USA.
[12] Gerardo Ceballos, Paul R. Ehrlich, and Rodolfo Dirzo. 2017. Biological annihilation via the ongoing sixth mass extinction signaled by vertebrate population losses and declines. In *PNAS,* July 25, 114 (30), E6089-E6096. DOI: https://doi.org/10.1073/pnas.1704949114
[13] Helene Cixous. 1976. The Laugh of the Medusa, trans. Keith Cohen and Paula Cohen. *Signs*, 1(4): 875-893.
[14] Alan Cooper, Robert Reimann, and Dave Cronin (eds.) 2007. *About Face 3. The essentials of interaction design*. Wiley, Indianapolis Ind, USA.
[15] Dan Goodley, Rebecca Lawthom, Kirsty Liddiard and Katherine Runswick-Cole. 2018. Posthuman Disability and DisHuman Studies. In *Posthuman Glossary*, Rosi Braidotti and Maria Hlavajova (eds.). Bloomsbury Academic, New York, US, London, UK. 342-345.
[16] Donna Haraway. 1988. Situated Knowledges: The Science Question in Feminism and the Privilege of Partial Perspective. *Feminist Studies* 14(3): 575–599.

[17] Donna Haraway. 1991 [1986]. A Cyborg Manifesto: Science, Technology, and Socialist-Feminism in the Late Twentieth Century. In *Simians, Cyborgs and Women: Reinvention of Nature*, pp. 149-181. Routledge, New York, USA.
[18] Donna Haraway. 1997. *Modest_witness@second_millenium.FemaleMan[©]_meets_OncoMouse™: feminism and technoscience.* Routledge, London, UK.
[19] Julia Kristeva. 1982. *Powers of Horror: An Essay on Abjection.* Columbia University Press, New York, USA.
[20] Julia Kristeva. 1989. *Tales of Love.* Columbia University Press, New York, USA.
[21] Ruth Levitas. 2013. *Utopia as Method.* Palgrave MacMillan, New York, USA.
[22] Genevieve Lloyd. 1984. *The Man of Reason: 'Male' and 'Female' in Western Philosophy*. Methuen Publishing, UK.
[23] Peter Markie. 2017. Rationalism vs. Empiricism. In *The Stanford Encyclopedia of Philosophy* (Fall 2017 Edition), Edward N. Zalta (ed.), URL: https://plato.stanford.edu/archives/fall2017/entries/rationalism-empiricism/.
[24] Richard Menary. 2010. Introduction to the special issue on 4E cognition. *Phenom Cogn Sci* 9: 459. DOI: https://doi.org/10.1007/s11097-010-9187-6.
[25] Mary Beth Rosson and John M. Carroll. 2002. Scenario-Based Design. In Julie A. Jacko and Andrew Sears (eds.), *The Human-Computer Interaction Handbook: Fundamentals, Evolving Technologies and Emerging Applications*, pp. 1032-1050.
[26] Londa Schiebinger. 1989. *The Mind Has No Sex? Women in the Origins of Modern Science.* Harvard University Press, Cambridge, MA, USA.
[27] Londa Schiebinger. 1993. *Nature's Body: Gender in the Making of Modern Science.* Beacon Press, Boston, USA.
[28] Cornelia Sollfrank (ed.). 2018. *Die schönen Kriegerinen: Technofeministische Praxis im 21. Jarhundert.* transversal texts, Vienna, Austria.
[29] Lucy Suchman. 2002. Figuring Service in Discourses of ICT: The Case of Software Agents. In *Proceedings of the IFIP TC8/WG8.2 Working Conference on Global and Organizational Discourse about Information Technology.* Kluwer, B.V. Deventer, The Netherlands. pp. 33-43.
[30] Shannon Winnubst. 2018. Decolonial Critique. In *Posthuman Glossary*, Rosi Braidotti and Maria Hlavajova (eds.). Bloomsbury Academic, New York, US, London, UK. 97-99.
[31] Dan Zahavi (ed.). 2012. *The Oxford Handbook of Contemporary Phenomenology.* Oxford University Press, Oxford, UK. DOI: 10.1093/oxfordhb/9780199594900.001.0001.